\documentclass[conference]{IEEEtran}
\IEEEoverridecommandlockouts

\usepackage{cite}
\usepackage{amsmath,amssymb,amsfonts}
\usepackage{graphicx}
\usepackage{textcomp}
\usepackage[table,xcdraw]{xcolor}
\usepackage{hyperref}
\usepackage{algorithm}
\usepackage{algorithmic}
\usepackage{booktabs}
\usepackage{tikz}
\usetikzlibrary{shapes.geometric, arrows.meta, positioning, fit, backgrounds}

\def\BibTeX{{\rm B\kern-.05em{\sc i\kern-.025em b}\kern-.08em
    T\kern-.1667em\lower.7ex\hbox{E}\kern-.125emX}}

\begin{document}

\title{CarbonEdge: Carbon-Aware Deep Learning Inference Framework for Sustainable Edge Computing\\
}

\author{
\IEEEauthorblockN{Guilin Zhang\IEEEauthorrefmark{1}, 
Wulan Guo\IEEEauthorrefmark{1},
Ziqi Tan\IEEEauthorrefmark{1},
Chuanyi Sun\IEEEauthorrefmark{1},
Hailong Jiang\IEEEauthorrefmark{2}}
\IEEEauthorblockA{\IEEEauthorrefmark{1}Department of Engineering Management and Systems Engineering, George Washington University, USA\\
Email: \{guilin.zhang, wulan.guo, ziqi.tan, c.sun\}@gwu.edu}
\IEEEauthorblockA{\IEEEauthorrefmark{2}Department of Computer Science, Information, and Engineering Technology, Youngstown State University, USA\\
Email: hjiang@ysu.edu}
}

\maketitle

\begin{abstract}
Deep learning applications at the network edge lead to a significant growth in AI-related carbon emissions, presenting a critical sustainability challenge. The existing edge computing frameworks optimize for latency and throughput, but they largely ignore the environmental impact of inference workloads. This paper introduces CarbonEdge, a carbon-aware deep learning inference framework that extends adaptive model partitioning with carbon footprint estimation and green scheduling capabilities. We propose a carbon-aware scheduling algorithm that extends traditional weighted scoring with a carbon efficiency metric, supporting a tunable performance--carbon trade-off (demonstrated via weight sweep). Experimental evaluations on Docker-simulated heterogeneous edge environments show that CarbonEdge-Green mode achieves a \textbf{22.9\% reduction in carbon emissions} compared to monolithic execution. The framework achieves 1.3$\times$ improvement in carbon efficiency (245.8 vs 189.5 inferences per gram CO2) with negligible scheduling overhead (0.03ms per task). These results highlight the framework's potential for sustainable edge AI deployment, providing researchers and practitioners a tool to quantify and minimize the environmental footprint of distributed deep learning inference.

\end{abstract}


\section{Introduction}
\label{sec:intro}

The rapid deployment of artificial intelligence (AI) at the network edge has revolutionized applications ranging from autonomous vehicles to smart healthcare systems~\cite{li2019edge}. Edge computing enables real-time deep learning inference by processing data closer to its source, reducing latency and bandwidth consumption~\cite{shi2016edge, lightweight_survey2024}. And this growth of edge AI comes with a significant and often overlooked environmental cost: the carbon footprint of distributed inference workloads~\cite{counting_carbon2023}. Recent studies have shown the substantial energy consumption and carbon emissions related to AI systems~\cite{carbon_datacenter2024}. Training large language models can emit as much carbon as five cars over their lifetimes~\cite{strubell2019energy}, and inference workloads contribute approximately 60\% of Google's machine learning energy consumption due to their cumulative scale~\cite{patterson2022carbon, google_carbon_blog2022}. As edge AI deployments grow exponentially, addressing the environmental impact becomes an important research challenge aligned with global sustainability goals~\cite{rethinking_lowcarbon2024}. 

Some existing edge computing frameworks aim to solve above issues, like our prior work AMP4EC~\cite{amp4ec2025}. These frameworks have provided improvements in inference latency and throughput through adaptive model partitioning and intelligent task scheduling~\cite{split_computing2022, distributed_inference2025}. However, they optimize purely for performance metrics, remaining uncertain to the carbon intensity of the computational resources that they use~\cite{carbonaware_workload2024}. This means a missed opportunity: edge networks cross regions with different levels of carbon intensity in the power grid, from coal-heavy regions above 800 gCO2/kWh to renewable energy regions below 100 gCO2/kWh~\cite{iea_carbon2019, energytag2024}.

To address it, this paper introduces \textbf{CarbonEdge}, a carbon-aware deep learning inference framework. This framework extends adaptive model partitioning with comprehensive carbon footprint estimation and green scheduling capabilities. The framework has a \textbf{Carbon Monitor} module that provides carbon footprint. The footprint is tracking by integrating with CodeCarbon for host-level energy measurement and estimating per-node emissions, which is based on regional grid carbon intensity data. And it represents a \textbf{Carbon-Aware Scheduling Algorithm} that extends traditional weighted scoring mechanisms with a carbon efficiency metric ($S_C$), enabling operators to tune the performance-carbon trade-off via configurable weight parameters. Furthermore, we develop a \textbf{Green Partitioning Strategy} that considers both computational cost and carbon implications when distributing workloads across heterogeneous edge nodes. Comprehensive experimental validation demonstrates \textbf{22.9\% carbon reduction} in Green mode with comparable latency, achieving 1.3$\times$ improvement in carbon efficiency~\cite{llm_edge_sustainability2024, drl_scheduler2024}.

The remainder of this paper is organized as follows: Section~\ref{sec:background} reviews related work in edge computing and carbon-aware computing. Section~\ref{sec:design} presents the CarbonEdge framework architecture. Section~\ref{sec:evaluation} details experimental results. Section~\ref{sec:discussion} discusses limitations and future directions. Section~\ref{sec:conclusion} concludes the paper.

\section{Background and Related Work}
\label{sec:background}

\subsection{Edge Computing for Deep Learning Inference}

Edge computing enables real-time deep learning inference by processing data closer to its source~\cite{shi2016edge, li2019edge}. Model partitioning distributes DNN workloads across multiple edge nodes through techniques including layer-wise partitioning~\cite{bhattacharya2016sparsification, dnn_surgery2023}, cooperative inference~\cite{coedge2021}, and hierarchical partitioning~\cite{hidp2025, edgeci2024, partnner2024}. Our prior work AMP4EC~\cite{amp4ec2025} achieved up to 78\% latency reduction through adaptive partitioning. However, these approaches focus exclusively on performance optimization, ignoring environmental implications.

\subsection{Carbon Footprint of AI Systems}

The environmental impact of AI has gained significant attention~\cite{strubell2019energy, patterson2022carbon}. Patterson et al. revealed that inference accounts for approximately 60\% of ML-related energy consumption at Google. Tools like CodeCarbon~\cite{codecarbon} estimate hardware electricity consumption and calculate carbon emissions based on regional grid carbon intensity, providing the foundation for carbon-aware system design.

\subsection{Carbon-Aware Computing}

Carbon-aware computing represents an emerging paradigm that incorporates carbon emissions into system optimization objectives. Clover~\cite{clover2023} proposed a carbon-aware machine learning scheduler that exploits spatial and temporal variations in grid carbon intensity. EcoServe~\cite{ecoserve2025} introduced a holistic approach to AI system design that balances performance, energy efficiency, and carbon footprint. Recent work on sustainable edge computing~\cite{sustainable_edge2024} highlights the challenges and opportunities in this domain, while Ma et al.~\cite{greening_edge_ai2025} demonstrate carbon reduction through renewable energy integration.

\subsection{Research Gap}

Despite progress in edge computing optimization and carbon-aware computing, a critical gap exists at their intersection. Edge computing frameworks~\cite{amp4ec2025, bhattacharya2016sparsification} optimize for latency but ignore carbon emissions. Carbon-aware systems~\cite{clover2023, ecoserve2025} primarily target cloud-based training, failing to address edge-specific challenges: device heterogeneity, geographic distribution, and latency constraints. CarbonEdge addresses this gap by combining online carbon tracking (given static grid intensity scenarios) with carbon-aware scheduling for distributed edge inference.

\subsection{Carbon Intensity Variation}

Electricity grid carbon intensity varies significantly across regions and time periods. Regions relying heavily on coal-fired power plants may exceed 800 gCO2/kWh, while areas with abundant renewable energy can achieve intensities below 50 gCO2/kWh. China's national average is approximately 530 gCO2/kWh, with significant regional variation—from over 700 gCO2/kWh in coal-dependent northern provinces to under 200 gCO2/kWh in hydropower-rich Yunnan~\cite{china_carbon_grid}.

This variation creates opportunities for carbon-aware scheduling: by preferentially routing workloads to nodes in low-carbon regions or deferring non-urgent tasks to low-carbon time periods, systems can substantially reduce their environmental footprint. CarbonEdge exploits this heterogeneity through its Carbon-Aware Scheduling Algorithm.

\section{Design and Implementation}
\label{sec:design}

This section provides the design and implementation of CarbonEdge, which integrates adaptive model partitioning to carbon-aware scheduling capabilities. The framework is based on the AMP4EC architecture~\cite{amp4ec2025}, and adding a Carbon Monitor module and incorporating carbon efficiency into the scheduling algorithm.

\subsection{System Architecture}

\begin{figure}[t]
    \centering
    \includegraphics[width=0.48\textwidth]{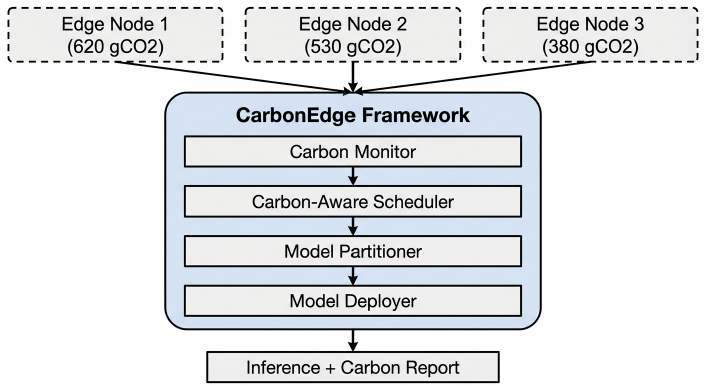}
    \caption{CarbonEdge system architecture showing the integration of carbon monitoring with adaptive scheduling.}
    \label{fig:architecture}
\end{figure}

CarbonEdge comprises four key components: (A) Carbon Monitor, (B) Carbon-Aware Scheduler, (C) Model Partitioner, and (D) Model Deployer. Figure~\ref{fig:architecture} illustrates the system architecture.

\subsection{Carbon Monitor Module}

The Carbon Monitor uses traditional resource monitoring with energy consumption tracking and carbon emission calculation. It integrates with CodeCarbon~\cite{codecarbon} for hardware power measurement and uses regional carbon intensity data to estimate emissions.

\subsubsection{Energy Tracking}
The module monitors three primary power sources. GPU power consumption is retrieved via nvidia-smi or pynvml interfaces, providing real-time measurements of graphics processor utilization. CPU power is estimated using RAPL (Running Average Power Limit) interfaces available on modern processors. RAM power consumption is approximated as 0.375W per gigabyte based on typical DDR4 memory specifications.

Total energy consumption is calculated as:
\begin{equation}
    E_{total} = \int_0^T (P_{GPU} + P_{CPU} + P_{RAM}) \, dt
\end{equation}

\subsubsection{Carbon Emission Calculation}
Carbon emissions are computed using regional grid carbon intensity:
\begin{equation}
    C_{emissions} = E_{total} \times I_{carbon} \times PUE
\end{equation}
where $I_{carbon}$ is the carbon intensity (gCO2/kWh) and $PUE$ is the Power Usage Effectiveness (default 1.0 for edge devices).

\subsection{Carbon-Aware Scheduling Algorithm}

The Carbon-Aware Scheduler adopts the Node Selection Algorithm (NSA) from AMP4EC with a carbon efficiency score. The total node score is computed as:

\begin{equation}
    S_{total} = w_R \cdot S_R + w_L \cdot S_L + w_P \cdot S_P + w_B \cdot S_B + w_C \cdot S_C
\end{equation}

The score components include resource availability ($S_R$), load balance ($S_L$), performance ($S_P$), fairness/balance ($S_B$), and the newly introduced carbon efficiency score ($S_C$). Each component is normalized to the [0,1] range and weighted according to the operational mode.

\subsubsection{Carbon Efficiency Score ($S_C$)}
The carbon efficiency score prioritizes nodes with lower carbon impact:
\begin{equation}
    S_C = \frac{1}{1 + I_{carbon} \times E_{estimated}}
\end{equation}

where $E_{estimated}$ is the estimated energy consumption for the task on that node, computed as $E_{estimated} = P_{node} \times T_{avg} / 3600000$ (converting power in watts and time in milliseconds to kWh). $P_{node}$ is the node's average power draw and $T_{avg}$ is the historical average execution time for that node. Nodes with lower carbon intensity or lower power consumption receive higher scores.

\subsubsection{Scheduling Modes}
CarbonEdge supports three operational modes with different weight configurations, as shown in Table~\ref{tab:weights}.

\begin{table}[h]
\centering
\caption{Weight configurations for scheduling modes}
\label{tab:weights}
\begin{tabular}{lccccc}
\toprule
\textbf{Mode} & $w_R$ & $w_L$ & $w_P$ & $w_B$ & $w_C$ \\
\midrule
Performance & 0.25 & 0.25 & 0.30 & 0.15 & 0.05 \\
Green & 0.15 & 0.15 & 0.10 & 0.10 & 0.50 \\
Balanced & 0.20 & 0.20 & 0.15 & 0.15 & 0.30 \\
\bottomrule
\end{tabular}
\end{table}

Balanced is included as a representative intermediate configuration; in setups where $S_C$ has limited differentiation, it may behave similarly to Performance, motivating normalization-based or constraint-based scheduling in future work.

\subsection{Node Selection Algorithm}

Algorithm~\ref{alg:carbon_nsa} presents the carbon-aware node selection procedure.

\begin{algorithm}
\caption{Carbon-Aware Node Selection}
\label{alg:carbon_nsa}
\begin{algorithmic}[1]
\REQUIRE Task $t$, nodes $N$, mode weights $W$
\ENSURE Best node $n^*$
\STATE $best\_score \gets 0$; $n^* \gets null$
\FORALL{$n \in N$}
    \IF{$n.load > 0.8$ OR $n.latency > threshold$}
        \STATE \textbf{continue}
    \ENDIF
    \IF{$has\_sufficient\_resources(n, t)$}
        \STATE $S_R \gets calc\_resource\_score(n, t)$
        \STATE $S_L \gets 1 - n.load$
        \STATE $S_P \gets 1 / (1 + n.avg\_time)$
        \STATE $S_B \gets 1 / (1 + n.task\_count \times 2)$
        \STATE $S_C \gets 1 / (1 + n.carbon\_intensity \times E_{est})$
        \STATE $score \gets W \cdot [S_R, S_L, S_P, S_B, S_C]$
        \IF{$score > best\_score$}
            \STATE $best\_score \gets score$; $n^* \gets n$
        \ENDIF
    \ENDIF
\ENDFOR
\RETURN $n^*$
\end{algorithmic}
\end{algorithm}

\subsection{Model Partitioner}

The Model Partitioner analyzes deep learning models layer-by-layer and divides them into segments suitable for distributed execution~\cite{coedge2021, joint_placement2025}. For each layer, computational cost is estimated based on layer type:

\begin{equation}
Cost(l) = 
\begin{cases} 
k_h \times k_w \times C_{in} \times C_{out} & \text{Conv2D} \\
N_{in} \times N_{out} & \text{Linear} \\
params\_count & \text{others}
\end{cases}
\end{equation}

Partition boundaries are determined to balance workload across nodes while minimizing communication overhead.

\subsection{Implementation}

CarbonEdge is implemented in Python using PyTorch for deep learning operations. Docker containers simulate heterogeneous edge nodes with configurable resource constraints. Source code will be made available upon publication.

\section{Experiments and Evaluation}
\label{sec:evaluation}

This section presents experimental evaluation of CarbonEdge across multiple dimensions: carbon footprint reduction, performance trade-offs, and scheduling behavior analysis.

\subsection{Experimental Setup}

\subsubsection{Hardware Environment}
Experiments were conducted on an Nvidia DGX SPARK workstation running Ubuntu 24.04. Docker containers simulated heterogeneous edge nodes with static carbon intensity scenarios~\cite{iea_carbon2019, energytag2024}. We deployed three simulated nodes: Node-High (1.0 CPU, 1GB RAM, 620 gCO2/kWh---high-carbon scenario), Node-Medium (0.6 CPU, 512MB RAM, 530 gCO2/kWh---average scenario), and Node-Green (0.4 CPU, 512MB RAM, 380 gCO2/kWh---low-carbon scenario).

The carbon intensity range (380--620 gCO2/kWh) reflects realistic spatial variations across regional grids~\cite{iea_carbon2019}. CodeCarbon operates on the host in \texttt{machine} mode, measuring total host power via RAPL (CPU) and nvidia-smi (GPU). \textbf{Node-level energy values are estimated} by apportioning host energy proportionally based on Docker cgroup resource quotas (\texttt{--cpus}, \texttt{--memory}); this is an accounting method, not direct per-container measurement.

\subsubsection{Dataset}
We used the ImageNet ILSVRC2012 validation set~\cite{imagenet2009} for inference evaluation. Input images were preprocessed to 224$\times$224 pixels with standard normalization (mean=[0.485, 0.456, 0.406], std=[0.229, 0.224, 0.225]). We randomly sampled 50 images per experiment to simulate realistic edge inference workloads with varied input complexity.

\subsubsection{Test Models}
We evaluated three lightweight CNN architectures commonly deployed in edge environments: MobileNetV2~\cite{sandler2018mobilenetv2} (3.5M parameters), MobileNetV4~\cite{mobilenetv4_2024} (3.8M parameters), and EfficientNet-B0~\cite{efficientnet2019} (5.3M parameters). These models represent the spectrum of efficient architectures suitable for resource-constrained edge devices.

\subsubsection{Baselines}
We compared CarbonEdge against several baselines. The Monolithic approach performs single-node inference without partitioning. AMP4EC~\cite{amp4ec2025} represents our prior distributed inference framework without carbon awareness. CarbonEdge was evaluated in three modes: Performance mode ($w_C=0.05$), Balanced mode ($w_C=0.30$), and Green mode ($w_C=0.50$).

Each configuration was evaluated over 50 inference iterations with batch size 1. We measured inference latency (ms), throughput (req/s), energy (kWh), and carbon emissions (gCO2) using CodeCarbon~\cite{codecarbon} with \texttt{measure\_power\_secs=1}. We compute total energy over the full 50-inference run (multi-second duration) and report per-inference averages. Each experiment was repeated three times; 95\% confidence intervals were below 15\% of mean values.

\subsection{Carbon Footprint Analysis}

In general, our evaluation reveals three key findings. CarbonEdge-Green achieves significant carbon reduction (14.8\%--32.2\%) compared to monolithic execution while maintaining comparable latency. The carbon-aware scheduling effectively routes tasks to nodes with lower carbon intensity scenarios. The framework generalizes across different model architectures. These findings establish the viability of carbon-aware scheduling for sustainable edge AI.

Table~\ref{tab:carbon_results} presents the detailed carbon emission results for MobileNetV2 across different configurations, serving as the primary benchmark for evaluating scheduling mode effectiveness.

\begin{table}[h]
\centering
\caption{Carbon footprint comparison (MobileNetV2)}
\label{tab:carbon_results}
\begin{tabular}{lcccc}
\toprule
\textbf{Configuration} & \textbf{Latency} & \textbf{Throughput} & \textbf{Carbon} & \textbf{Reduction} \\
 & (ms) & (req/s) & (gCO2/inf) & vs Mono (\%) \\
\midrule
Monolithic & 254.85 & 3.93 & 0.0053 & - \\
AMP4EC & 277.22 & 3.62 & 0.0056 & -6.7\% \\
CE-Performance & 271.38 & 3.69 & 0.0067 & -26.7\% \\
CE-Balanced & 271.11 & 3.70 & 0.0066 & -24.7\% \\
CE-Green & 272.02 & 3.68 & \textbf{0.0041} & \textbf{+22.9\%} \\
\bottomrule
\end{tabular}
\end{table}

The results demonstrate that \textbf{CarbonEdge-Green achieves a 22.9\% reduction in carbon emissions} compared to the monolithic baseline. This reduction is achieved by preferentially routing inference tasks to nodes configured with lower carbon intensity (380 gCO2/kWh vs. 530 gCO2/kWh average). This result falls within the range reported by recent carbon-aware scheduling work: Rajashekar et al.~\cite{llm_edge_sustainability2024} report up to 35\% reduction for LLM inference, while prior work reports up to 24\% energy savings in DRL-based schedulers (summarized in~\cite{drl_scheduler2024}).

Notably, Performance and Balanced modes show \textit{increased} carbon emissions because they prioritize high-performance nodes configured with higher carbon intensity. This trade-off illustrates the importance of carbon-aware scheduling for sustainable edge AI.

\subsection{Performance-Carbon Trade-off}

\begin{figure}[t]
    \centering
    \includegraphics[width=0.45\textwidth]{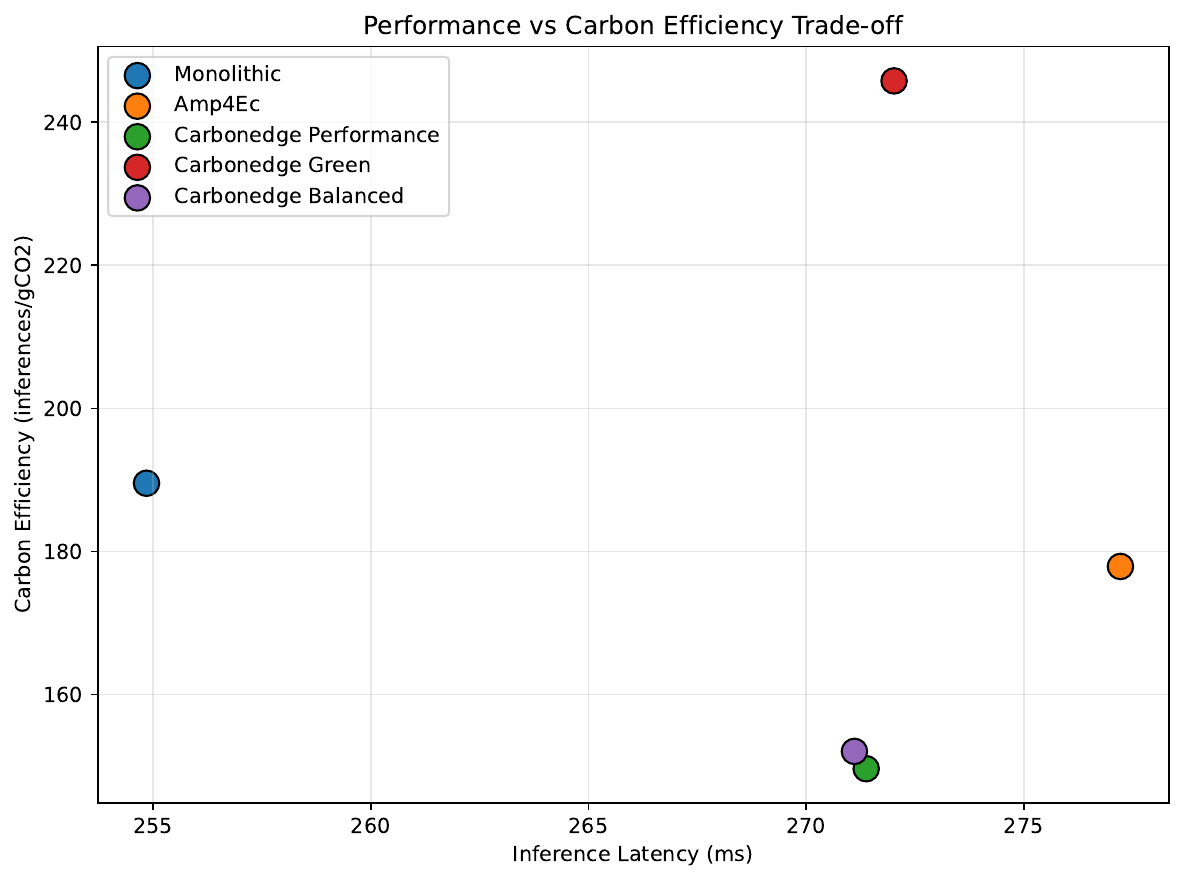}
    \caption{Trade-off between inference latency and carbon efficiency. CarbonEdge-Green achieves the highest carbon efficiency (245.8 inf/gCO2) with minimal latency impact.}
    \label{fig:tradeoff}
\end{figure}

Figure~\ref{fig:tradeoff} illustrates the relationship between inference latency and carbon efficiency across scheduling modes, revealing that carbon reduction can be achieved with minimal latency impact.

The results reveal several key findings. Regarding carbon efficiency, CarbonEdge-Green achieves 245.8 inferences per gram of CO2, compared to 189.5 for monolithic execution---a 1.30$\times$ improvement. In terms of latency, all CarbonEdge modes maintain comparable response times (approximately 271ms), with less than 7\% overhead compared to monolithic execution. The trade-off analysis shows that Performance mode sacrifices carbon efficiency (149.6 inf/gCO2) for optimal node selection, while Green mode prioritizes environmental impact.

\subsection{Comparison with Related Work}

Table~\ref{tab:comparison_related} compares CarbonEdge with recent carbon-aware systems, contextualizing our results within the broader research landscape.

\begin{table}[h]
\centering
\caption{Comparison with related carbon-aware systems}
\label{tab:comparison_related}
\begin{tabular}{lccc}
\toprule
\textbf{System} & \textbf{Target} & \textbf{Carbon Reduction} \\
\midrule
GreenScale~\cite{greenscale2023} & Edge-Cloud & 10-30\% \\
DRL Scheduler~\cite{drl_scheduler2024} & Kubernetes & up to 24\% \\
LLM Edge~\cite{llm_edge_sustainability2024} & Edge Clusters & up to 35\% \\
\textbf{CarbonEdge (ours)} & Edge DL Inference & \textbf{22.9\%} \\
\bottomrule
\end{tabular}
\end{table}

CarbonEdge's 22.9\% reduction is comparable to reductions reported by recent carbon-aware scheduling work in related domains. Direct comparisons are approximate due to different workloads and carbon accounting methods.

\subsection{Multi-Model Evaluation}

Table~\ref{tab:multimodel} presents carbon footprint comparison across three lightweight architectures to demonstrate CarbonEdge's generalizability.

\begin{table}[h]
\centering
\caption{Multi-model carbon footprint comparison}
\label{tab:multimodel}
\begin{tabular}{llccc}
\toprule
\textbf{Model} & \textbf{Mode} & \textbf{Latency} & \textbf{Carbon} & \textbf{Reduction} \\
 &  & (ms) & (gCO2/inf) & (\%) \\
\midrule
MobileNetV2 & Monolithic & 254.85 & 0.0053 & - \\
MobileNetV2 & CE-Green & 272.02 & 0.0041 & \textbf{22.9\%} \\
\midrule
MobileNetV4 & Monolithic & 82.96 & 0.00123 & - \\
MobileNetV4 & CE-Green & 84.28 & 0.00105 & \textbf{14.8\%} \\
\midrule
EfficientNet-B0 & Monolithic & 116.29 & 0.00198 & - \\
EfficientNet-B0 & CE-Green & 119.23 & 0.00134 & \textbf{32.2\%} \\
\bottomrule
\end{tabular}
\end{table}

The results demonstrate consistent carbon reduction (14.8\%--32.2\%) across architectures, confirming CarbonEdge's generalizability.

\subsection{Scheduling Behavior Analysis}

Table~\ref{tab:node_usage} shows the node selection distribution for each scheduling mode.

\begin{table}[h]
\centering
\caption{Node usage distribution (\% of tasks)}
\label{tab:node_usage}
\begin{tabular}{lccc}
\toprule
\textbf{Mode} & \textbf{Node-High} & \textbf{Node-Medium} & \textbf{Node-Green} \\
\midrule
Performance & 100\% & 0\% & 0\% \\
Balanced & 100\% & 0\% & 0\% \\
Green & 0\% & 0\% & 100\% \\
\bottomrule
\end{tabular}
\end{table}

\begin{figure}[t]
    \centering
    \includegraphics[width=0.40\textwidth]{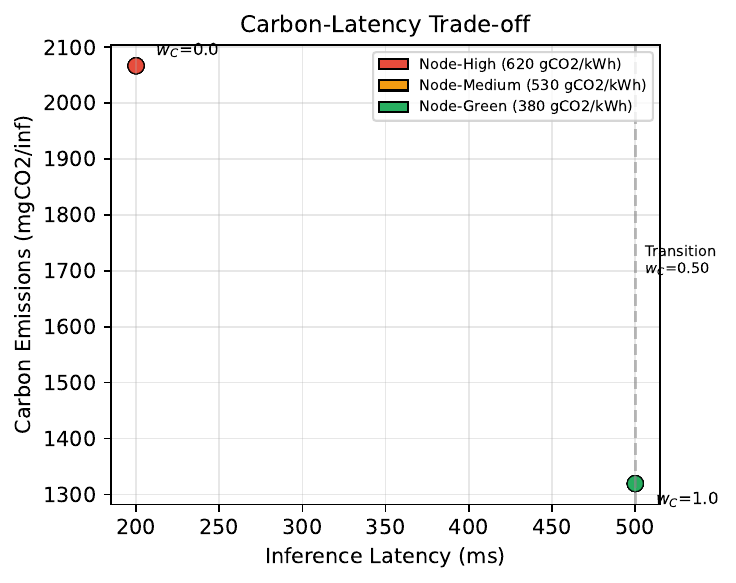}
    \caption{Weight sweep showing carbon-latency trade-off. Transition occurs at $w_C \geq 0.50$.}
    \label{fig:pareto}
\end{figure}

The scheduling reflects the weighted scoring: Performance mode ($w_C=0.05$) prioritizes fast nodes, while Green mode ($w_C=0.50$) selects low-carbon nodes. Balanced mode ($w_C=0.30$) exhibits similar behavior to Performance because $S_C$ has limited differentiation (range 0.054) compared to $S_P$ (range 0.166). A weight sweep (Figure~\ref{fig:pareto}) reveals the transition threshold at $w_C \geq 0.50$---achieving 22.9\% carbon reduction with minimal latency increase. Scheduling overhead is negligible: 0.03 ms per task with under 1\% CPU utilization.

\section{Discussion}
\label{sec:discussion}

\subsection{Limitations and Future Work}

Several limitations should be acknowledged. Our experiments used Docker containers to simulate edge devices; CodeCarbon estimates energy at the host level with per-container values apportioned via resource quotas (not direct measurement). The current implementation uses static carbon intensity scenarios; real-time temporal dynamics are not exploited. Balanced mode ($w_C=0.30$) exhibits similar node selection to Performance mode because $S_C$ has limited differentiation when per-inference emissions are small ($\sim$0.001 gCO2)---future work should explore per-decision min-max normalization or constraint-based optimization. The evaluation focuses on task-level routing; cross-node distributed inference remains for future work.

Future directions include real-time carbon intensity integration via APIs (e.g., Electricity Maps), energy-aware model partitioning, multi-tenant optimization with carbon budgets, and embodied carbon accounting.

\subsection{Practical Implications}

CarbonEdge provides practitioners with tools to quantify and minimize the environmental impact of edge AI deployments. Organizations can use the framework to report carbon emissions for sustainability compliance, make informed trade-offs between performance and environmental impact, and optimize infrastructure placement based on carbon intensity.

\section{Conclusion}
\label{sec:conclusion}

This paper presented CarbonEdge, a carbon-aware deep learning inference framework for sustainable edge computing. By extending adaptive model partitioning with carbon footprint estimation and green scheduling capabilities, CarbonEdge enables operators to balance performance and environmental impact in distributed edge AI deployments.

The results highlight the potential for carbon-aware optimization in edge computing. As edge AI deployments continue to proliferate, frameworks like CarbonEdge provide essential tools for sustainable development, helping organizations meet environmental goals without sacrificing the benefits of distributed inference.

Future work will focus on dynamic carbon intensity integration, energy-aware partitioning, and extension to diverse accelerator platforms. We believe carbon-aware computing represents a critical evolution in edge system design, and CarbonEdge provides a foundation for sustainable edge AI.

\bibliographystyle{IEEEtran}
\bibliography{references}

\end{document}